\documentclass[pra,twocolumn,a4paper,showpacs]{revtex4-1}
\usepackage{amsmath, amsthm, amsfonts, amssymb}
\usepackage{graphicx}
\usepackage{subfig}
\usepackage{natbib}
\usepackage{epstopdf}
\usepackage{caption}
\usepackage{footnote}
\input amssym.def
\usepackage{url}
\begin{document}

\title{Bose-Einstein Condensation in an electro-pneumatically transformed quadrupole-Ioffe magnetic trap}
\author{Sunil Kumar, Sumit Sarkar, Gunjan Verma, Chetan Vishwakarma, Md. Noaman{$^1$}
 and Umakant Rapol}
\email {Author to whom correspondence should be addressed. Electronic mail: umakant.rapol@iiserpune.ac.in}

\affiliation{Department of Physics, Indian Institute of Science Education and Research, Pune 411008, Maharashtra, India}
\begin{abstract}
We report a novel approach for preparing a Bose-Einstein condensate (BEC) of $^{87}$Rb atoms using electro-pneumatically driven transfer of atoms into a Quadrupole-Ioffe magnetic trap (QUIC Trap). More than 5$\times$$10^{8}$ atoms from a Magneto-optical trap are loaded into a spherical quadrupole trap and then these atoms are transferred into an Ioffe trap by moving the Ioffe coil towards the center of the quadrupole coil, thereby, changing the distance between quadrupole trap center and the Ioffe coil. The transfer efficiency is more than 80 \%. This approach is different from a conventional approach of loading the atoms into a QUIC trap wherein the spherical quadrupole trap is transformed into a QUIC trap by changing the currents in the quadrupole and the Ioffe coils. The phase space density is then increased by forced rf evaporative cooling to achieve the Bose-Einstein condensation having more than $10^{5}$ atoms. 
\end{abstract}
\maketitle
The first experimental demonstration of BEC in alkali atoms \cite{anderson1995observation, PhysRevLett.79.1170, PhysRevLett.75.3969} led to a rapid growth of research in the area of ultracold quantum gases. It has given a completely new insight into the study of quantum matter at ultralow temperatures and opened up new experimental test-bed to study condensed matter systems. It has provided the ability to emulate real-life condensed matter systems to be able to gain new insights into supercoductivity and superfluidity. BECs subjected to periodic \cite{bloch2005ultracold,morsch2006dynamics} and disordered optical potentials \cite {PhysRevA.81.063639,PhysRevLett.95.250403,PhysRevLett.95.170411} are some such examples of quantum emulators. In addition, extreme control through light-matter interaction has enabled creation and study of artificial gauge
potentials \cite{RevModPhys.83.1523, galitski2013spin} as a path towards understanding new materials like topological insulators and simulation of lattice gauge theories in high energy physics \cite{PhysRevLett.110.125304, RevModPhys.55.775}.

Development of robust and yet simple experimental system for routine production of BECs is always a challenge during the design phase. It often involves a tradeoff between optical, mechanical and control system complexities and modularity while being able to achieve the desired scientific goals. The final challenge is the design of the trap. Driven again by the scientific goals, one chooses between a magnetic trap \cite{PhysRevLett.77.416}, all-optical trap \cite{barrett2001all} or a combined optical-magnetic trap \cite{PhysRevA.79.063631}. Historically, the magnetic trap with different variants has been the first of the most widely used traps. 

The simplest example of creating a magnetic trap is to use spherical quadrupole trap using a pair of coils in anti Helmholtz configuration, in which magnetic field crosses zero at the center and increases linearly with the distance from the trap center. The quadrupole trap offers tightest confinement at the cost of loss of cold atoms from the trap due to non adiabatic spin flips \cite{PhysRevLett.75.3969, PhysRevLett.74.3352} at the trap center, known as Majorana spin flips which prohibits increase of phase space density and thus the formation of BEC. There are several techniques for creating non zero minima such as time-averaged orbiting potential (TOP) trap \cite{PhysRevLett.74.3352}, optical dipole trap \cite{PhysRevA.71.011602} and Ioffe-Pritchard (IP) type trap \cite{PhysRevA.58.R2664,PhysRevLett.77.416,PhysRevLett.51.1336}. IP type traps can give very high axial and radial trapping frequencies giving rise to very tight confinement as compared to a TOP trap. Tight confinement results in very high densities and large collision rates which is very crucial during evaporative cooling \cite{PhysRevB.34.3476}, giving a thermalization time shorter than the lifetime of the atoms in magnetic trap. Although the IP type trap has been a successful trap that was widely used in the initial period after the demonstration of BEC, IP traps have certain disadvantages such as the mode matching of spatially separated center of Magneto-Optical Trap (MOT) with the center of magnetic trap, limited optical access to the trapping region and large power dissipation with relatively complex electronic control circuits for switching the magnetic trap which requires additional power supplies. The advent of a modified quadrupole-Ioffe trap \cite{PhysRevA.58.R2664}, solved the problem of mode matching and in addition had a simplified design consisting of only three coils (the two quadrupole coils and a third Ioffe coil having axis perpendicular to the axis of the quadrupole coils). This scheme involves independent control of currents through the quadrupole and Ioffe coil. In addition, the geometrical design of the Ioffe coil has to be such that it does not obstruct the MOT laser beams in some cases. This poses some limitation in getting the desired trapping frequencies.

In this report, we report a novel scheme of an Ioffe-Pritchard magnetic trap wherein, atoms trapped in a spherical-quadrupole magnetic trap are transferred into a quadrupole-Ioffe (QUIC) trap by mechanically translating the Ioffe coil. The transfer of magnetically trapped atoms over large distances was first demonstrated by Greiner {\it et. al.} \cite{PhysRevA.63.031401} by using a chain of quadrupole coils, in which the atoms were moved up to 330 mm and then quadrupole potential is converted to Ioffe type potential using QUIC trap geometry. Another method of creating a Ioffe type potential in which quadrupole coils are mounted on a linear actuator which is driven by a servo motor is given in Refs. \cite{lewandowski2003simplified,nakagawa2005simple}. The atoms are transported up to 550 mm to reach into the ultra high vacuum region to create IP type potential \cite{lewandowski2003simplified,nakagawa2005simple}. All these transfer mechanisms involve a sophisticated electronic switching circuit for controlling the currents in all the coils to avoid heating of trapped atom in magnetic trap during the transfer process.
In our trap, we have solved the problem by using a single UHV chamber in which MOT and magnetic trap are created in the same chamber. Hence, requires less transportation distance up to 40 mm only, as compared to previous works \cite{PhysRevA.63.031401,lewandowski2003simplified,nakagawa2005simple}. Another advantage in our set up is moving a much smaller Ioffe coil in comparison to moving the large quadrupole coils. In our trap the transfer mechanism is purely based on the transportation distance and it does not need a current control in separate coils, thus simplifying the electronic control circuit with the use of a single current controller and a single IGBT switch.

\begin{figure}[ht]
\centering
\includegraphics[width=0.75\linewidth, angle=270]{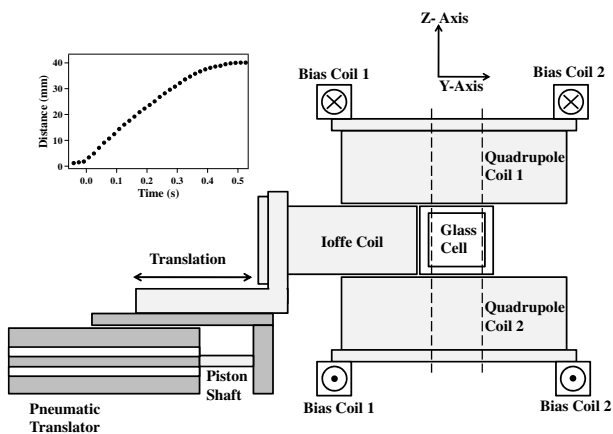} 
\captionsetup{singlelinecheck=off,justification=raggedright}
\caption{Schematic of the magnetic trap. The Ioffe coil is mounted on a pneumatic translator that moves along the y-axis. The axis of the quadrupole coils is along the gravity (z-axis). A pair of Helmholtz coils along the y-axis provide for the bias magnetic field to the QUIC trap. The glass cell has a cross-sectional dimensions of 39 mm $\times$ 39 mm and 150 mm length. The spacing between the quadrupole coils is 48 mm. The horizontal MOT beams are in the x-y plane at an angle of 45$^\circ$ to the length of the glass cell. The vertical MOT beam goes through the axis of the quadrupole coils along the z-axis. The Zeeman slowing beam propagates along the x-direction into the plane of the paper. Inset shows a typical position vs time profile of the Ioffe coil during translation. The velocity profile remains nearly linear in the central region of the translation. The diagram is not to scale. }
\end{figure}
The final configuration of the trap is the widely used QUIC trap as reported in Ref. \cite{PhysRevA.58.R2664}. By means of forced rf evaporative cooling, we report fast production of BEC in such a trap. The mechanical translation of the Ioffe coil is achieved by mounting the Ioffe coil on a pneumatically actuated translator as shown in Fig. 1. As mentioned earlier, this transport scheme provides two major advantages to other conventional schemes viz. a) It reduces the complexity of electronic control system. In our setup, we use a single DC power supply (Delta Elektronika Model SM60-100) for controlling and a single IGBT (Make EUPEC, model no BSM300GB120DLC) for switching the current through all the coils. 
b) Provides added flexibility in designing stiffer magnetic traps with improved optical access to laser cooling beams. The geometrical constraints posed by the size of our glass cell (39 mm $\times$ 39 mm) became critical in designing Ioffe coil with tolerable levels of electrical power consumption. Hence, we had to go for a non-conical shaped Ioffe coil. This non-conical Ioffe coil if present fixed, would obstruct the laser beams that intersect the glass cell at 45 degrees.
\begin{figure}[ht]
\centering
\includegraphics[width=\linewidth]{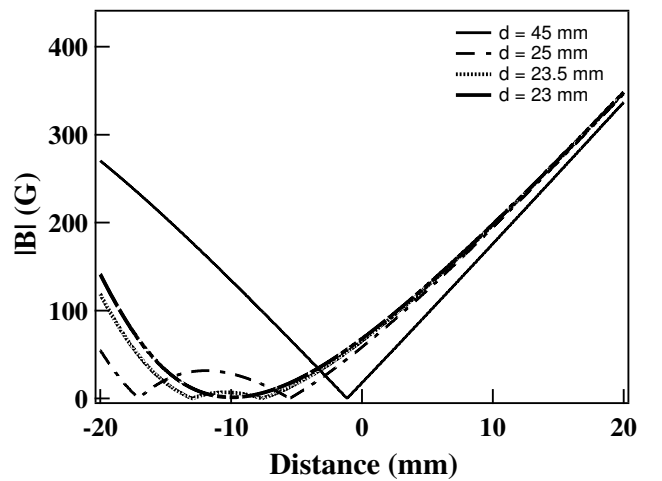} 
\captionsetup{singlelinecheck=off,justification=raggedright}
\caption{Absolute value of the magnetic field along the axis of Ioffe coil vs the distance of the Ioffe coil and the quadrupole trap center. The current through all the coils is 27A and the distance between spherical quadrupole trap centre and Ioffe coil is (a) 45 mm, (b) 25 mm, (c) 23.5 mm, (d) 23 mm. Note that the transformation of the linear potential with zero minima to a harmonic trap with non-zero minima is similar to that which can be achieved by varying currents through fixed Ioffe and quadrupole coils \cite{PhysRevA.58.R2664}.}
\end{figure}
The pneumatic translator is a commercially available product (Festo, Model No. SLT-16-40-P-A) that provides a total translation of 40 mm. The translator is a two position translator that is guaranteed to give a positioning repeatability of $\pm$20 $\mu$m. The position verses time profile of the translator is shown in inset Fig. 1. The applied air pressure was about 4 bar. The peak forward and reverse speed of the translator can be controlled by a pair of flow control valves. 

Each of the two anti-Helmholtz coils is made of 15 layers with 19 turns in each layer wound with 17 AWG enameled magnet wire. Each layer is separated by 1 mm thick spacers. The Ioffe coil has 7 layers with 20 turns in each layer. Each of the two bias coils is made by winding 15 turns of insulated copper tubing of 3 mm OD and 2 mm ID. The anti-Helmholtz coils produce a magnetic field gradient of 13 G/cm/A in the axial direction. The bias coils generate a constant field of 0.84 G/A. The distance between the bias coils is 180 mm. All the coils except the bias field generating coils are enclosed in water tight Delrin assemblies and cold water is circulated to remove heat generated during the operation of the trap. In Fig. 2, we show the numerical simulation of the magnitude of the magnetic field along the axis of the Ioffe coil for different distances of the Ioffe coil from the center of the quadrupole coils for a current of 25 A passing through all the coils simultaneously including the bias coils. When the Ioffe coil is about 45 mm away from the axis of the quadrupole coils, the effect of the magnetic field created by the Ioffe coil is negligible. As the Ioffe coil moves closer to the quadrupole coil axis, a double minima in the magnetic potential starts appearing and the two minima merge at a distance of 23 mm.

\begin{figure}[ht]
\centering
\includegraphics[width=\linewidth]{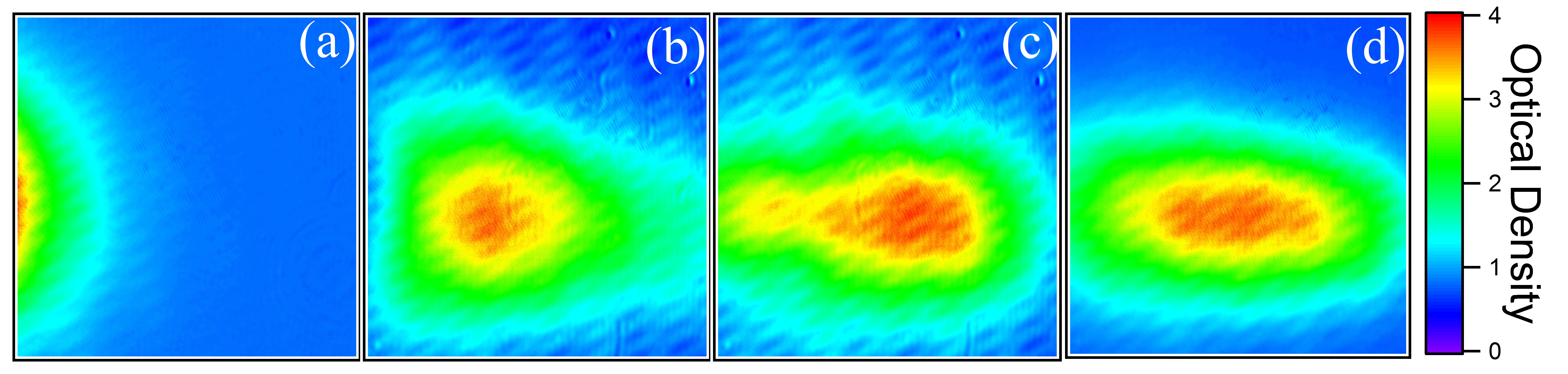} 
\captionsetup{singlelinecheck=off,justification=raggedright}
\caption{Series of images demonstrating transfer of atoms during the translation of the Ioffe coil. Images of the cloud are taken for distances between the quadrupole trap and Ioffe coil is (a) 45 mm (pure quadrupole trap. since, the position of the quadrupole trap minima is located outside the field of view of the camera, a majority of the portion of the cloud is not visible) (b) 25 mm (c) 23.5 mm (d) 23 mm. The field of view of the images is 8 mm $\times$ 8 mm}
\end{figure}
Our experimental setup consists of three sections viz. a) a Rb effusion source, b) a decreasing field Zeeman slower and c) a rectangular quartz cell. The Rb effusion source and Zeeman slower designs are adapted from \cite{PhysRevA.79.063631}. A 75 l/s ion pump is used to pump the Rb effusion source region, two 150 l/s ion pumps are used before and after the Zeeman slower for pumping. In addition, there are two Ti sublimator pumps along with the two 150 l/s pumps. There is a Non-Evaporable Getter (NEG) pump near the Quartz cell.
Atomic beam emanating from the Zeeman slower is used to load the MOT in the quartz cell. The MOT beams are orthogonal to each other and are incident at 45$^\circ$ to the cell. We load about 1 $\times$ 10$^9$ $^{87}$Rb atoms in about 35 s time. During the loading of the atoms in the MOT, the field gradient of the quadrupole coils is $\sim$13 G/cm and the detuning of the cooling beams is -2 $\Gamma$ from the 5$^2$S$_{1/2}$ $|F=2\rangle \longrightarrow$ 5$^2$P$_{3/2}$ $|F'=3\rangle$ state, where $\Gamma$ = 6.1 MHz is the linewidth of the excited state. The laser beams have a 1/$e^2$ diameter of 22 mm and the intensity of the laser light is about 3$\times$ the saturation intensity of the 5$^2$S$_{1/2}$ $|F=2\rangle \longrightarrow$ 5$^2$P$_{3/2}$ $|F'=3\rangle$ transition. As mentioned earlier, during the entire loading sequence of the atoms, the quadrupole coils, Ioffe coils and the bias coils are connected in series. It has been seen that there is no appreciable effect on the nature of loading of the atoms in the MOT when the Ioffe and bias coils are disconnected. This is due to the fact that, the shift in the center of the quadrupole field is less than 0.5 mm which is negligible in comparison to the size of the cloud ($\sim$ 8 mm). 
Atoms are then compressed in the compressed-MOT stage by increasing the detuning to -4 $\Gamma$ while reducing the intensity to 1/10$^{th}$ of the saturation intensity. Atoms are subjected to a polarization gradient cooling stage for 4 ms where the magnetic field is turned off while the detuning is ramped to -8 $\Gamma$ and intensity of the MOT beams is kept unchanged from the previous stage. At this stage we have about 8 $\times$ 10$^8$ atoms in the trap at a temperature of $\sim$30 $\mu$K. Atoms are then optically pumped into the 5$^2$S$_{1/2}$ $|F=2, m_f=2\rangle$ state by applying a 250 $\mu$s pulse of circularly polarized laser driving the 5$^2$S$_{1/2}$ $|F=2\rangle \longrightarrow$ 5$^2$P$_{3/2}$ $|F'=3\rangle$ transition in the presence of a small bias magnetic field in the z-direction.

\begin{figure}[htb]
\centering
\includegraphics[width=\linewidth]{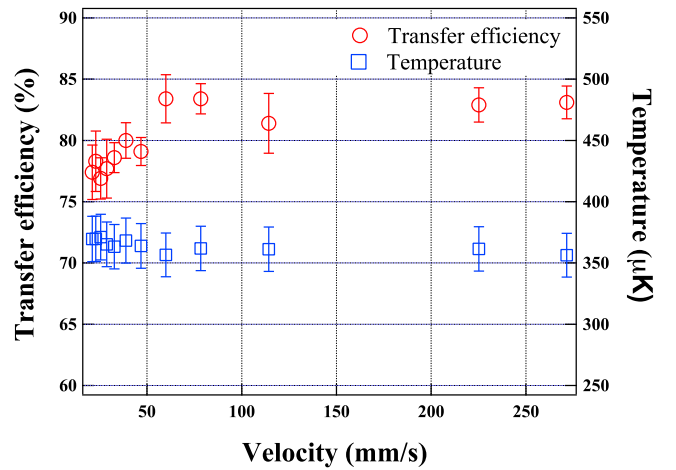} 
\captionsetup{singlelinecheck=off,justification=raggedright}
\caption{Transfer effeciency of atoms change in the temperature of the cloud from quadrupole to QUIC trap}
\end{figure}

By rapidly turning on the quadrupole magnetic field to 91 G/cm in less than 2 ms we capture more than 5 $\times$ 10$^8$ in the spherical quadrupole magnetic trap. The temperature in the magnetic trap at this stage is a little over 500 $\mu$K. Atoms are then compressed adiabatically in 1500 ms from 91 G/cm to 325 G/cm magnetic field gradient. After this stage the pneumatic translator is triggered to initiate the translation of the Ioffe coil. Within less than 1 s the Ioffe coil reaches its final position where it almost touches the quartz cell. As shown in Fig. 2, as the coil keeps moving towards the center of the quadrupole trap, at a certain distance there appear two minima in the magnetic trapping potential and then they finally merge when the coil reaches its end position. We have captured the transfer process of atoms in the QUIC trap by turning off the magnetic field at different stages of the position of the Ioffe coil. Fig. 3 shows the absorption images of the atoms during the transfer process. We transfer more than 80 $\%$ atoms from the spherical quadrupole trap into the Ioffe trap. A systematic study of the transfer efficiency of the atoms into the Ioffe trap as a function of the speed of translation of the Ioffe coil has been done. Data in the Fig. 4 shows that there is a gradual increase in the transfer efficiency (up to 85 $\%$) of the atoms for peak speeds up to 80 mm/s and then the efficiency drops a little. However, the range of efficiencies is limited to a narrow range between 75 $\%$ and 85$\%$. This experiment has been performed only in the available range of speeds that could be accessed by controlling the flow rate of the entrance valve of the Pneumatic translator while keeping a constant pressure of 4 bar. 
\begin{figure}[htb]
\centering
\includegraphics[width=\linewidth]{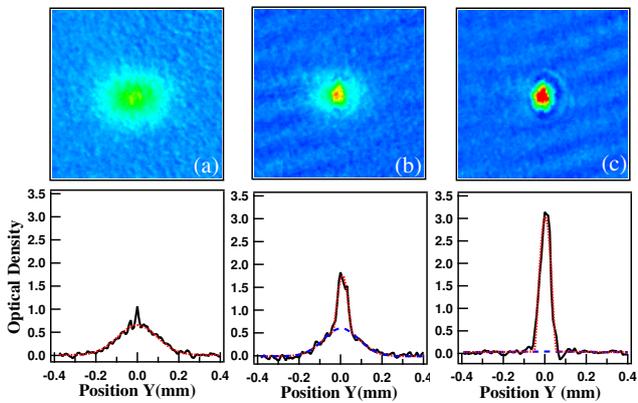} 
\captionsetup{singlelinecheck=off,justification=raggedright}
\caption{Typical images of the ultracold cloud of atoms taken after 21 ms of expansion under gravity (a) Thermal cloud at temperature about 211 nK (final rf frequency 1.94 MHz) (b) bimodal density distribution at temperature 171 nK (final rf frequency 1.93 MHz) showing the emergence of a sharp peak at the center with a condensate fraction of 23$\%$. (c) Almost pure condensate at final rf frequency 1.919 MHz, The condensate fraction is 97$\%$. Lower set of curves are the horizontal cross-section plots of the images above. The field of view of the images is 0.8 mm $\times$ 0.8 mm.}
\end{figure}
The rise in temperature also does not change appreciably to affect the forced evaporative cooling process in the later stages. The QUIC trap used in our setup has a radial frequency of 2$\pi$ $\times$ 140 Hz and axial frequency of 2$\pi$ $\times$ 21 Hz which have been measured by perturbing the cloud of cold atoms. The bias field is $\sim$1 G. The measured axial field gradient of the trap is 325 G/cm and the estimated axial field curvature is 196 G/cm$^2$. Rf evaporative cooling is performed to further cool the atoms in Ioffe trap by rf-induced spin flips. The rf frequency was swept from 40 MHz to 1.919 MHz over a time period of 16.4 s in different stages. After evaporation, thermalized cloud is probed by using absorption imaging system which consists of single lens geometry and EMCCD camera having a magnification factor of 1.0(1). Cooled atoms are released from the trap by switching off the magnetic trap. After the atoms expand ballistically, near-resonant laser pulse of 40 $\mu$s is illuminated on the expanding cloud and the shadow is imaged onto the camera. Atoms' number density, temperature and the total number were calculated by analyzing these absorption images. When rf frequency is lowered below 1.94 MHz a sudden appearance of bimodal distribution is observed in the time of flight images shown in Fig.5, which is the signature of BEC phase transition. The calculated number of atoms in the BEC is 3.5(4) $\times$ 10$^5$. The instability of the bottom of the trap is  below 3 mG measured over a period of couple of hours. Thereby, proving the usability of this design.

In conclusion, we have demonstrated a novel method for realization of Bose-Einstein condensation of rubidium atoms in electromechanical QUIC trap. By first loading into pure quadrupole trap, atoms are transferred into the pure Ioffe trap by changing the position of Ioffe coil. This type of trap reduces technical complexity associated with the electronics and delivers an added advantage of optimizing magnetic trap parameters. The design of this trap geometry allows full optical access during loading in the magnetic trap. In addition, prepared BEC can be loaded into additional optical traps and by moving the Ioffe away can provide additional optical access. Due to an added advantage of designing tighter magnetic traps, one can do a fast evaporation.

UR acknowledges the funding received from Indian Institute of Science Education and Research, Pune. SK would like to acknowledge Council of Scientific and Industrial Research (CSIR), India for research fellowship.

\bibliography{biblio}{}
\bibliographystyle{aapmrev4-1}
\end{document}